\input{aipcheck}

\documentclass[
    ,final            % use final for the camera ready runs
  ]
  {aipproc}

\layoutstyle{6x9}
\usepackage{txfonts}
\def\apj{Ap. J.}

\def\apjs{Ap. J. Supp.}
\def\apss{Astrophys. \& Space Sci.}

\def\physrep{Physics Reports}
\def\aap{Astron. \& Astrophys.}

\def\nat{Nature}

\begin{document}

\title{The GLAST mission, LAT and GRBs}


\author{Nicola Omodei}{
  address={INFN Pisa, Largo B.Pontecorvo, 3 56100 Pisa }
}

\author{GLAST/LAT GRB Science Group}{
  address={http://glast.gsfc.nasa.gov/science/grbst/members.html}
}

\begin{abstract}
The GLAST Large Area Telescope (LAT) is the next generation satellite experiment for high-energy gamma-ray astronomy.
It is a pair conversion telescope built with a plastic anticoincidence shield, a segmented CsI electromagnetic calorimeter, and the largest silicon strip tracker ever built. It will cover the energy range from 30 MeV to 300 GeV, shedding light on many issues left open by its predecessor EGRET.
One of the most exciting science topics is the detection and observation of gamma-ray bursts (GRBs).
In this paper we present the work done so far by the GRB LAT science group in studying the performance of the LAT detector to observe GRBs. We report on the simulation framework developed by the group as well as on the science tools dedicated to GRBs data analysis. We present the LAT sensitivity to GRBs obtained with such simulations, and, finally, the general scheme of GRBs detection that will be adopted on orbit.
\end{abstract}

\maketitle

%%%%%%%%%%%%%%%%%%%%%%%%%%%%%%%%%%%%%%%%%%%%
%% MAINMATTER
%%%%%%%%%%%%%%%%%%%%%%%%%%%%%%%%%%%%%%%%%%%%

\section{Introduction}

The Gamma-ray Large Area Space Telescope (GLAST) is an international mission that will
study the gamma-rays Universe\footnote{For more details, see the GLAST website at: \url{http://glast.gsfc.nasa.gov/}}.
GLAST, scheduled for launch in late-2007, is instrumented with a tracker of silicon strip planes with slabs of tungsten converter, followed by an hodoscopic calorimeter; the tracker, is an array of $4\times 4$ identical towers, surrounded by an anticoincidence detector (ACD) which identifies charged cosmic rays. This pair production telescope, called the Large Area Telescope (LAT), is sensitive to gamma rays in the energy range between 30~MeV-300~GeV and above.
The LAT's energy range, field-of-view (FoV) and angular resolution are vastly
improved in comparison with those  its highly-successful predecessor EGRET (1991-2000), so that the LAT will provide a factor 30 or more advance in sensitivity.
This improvement should enable the detection of several thousands of new high-energy sources and allow the study of gamma-ray bursts (GRBs) and other transients, the resolution of the extragalactic diffuse gamma-ray emission, the search for dark matter and the detection of active galactic nuclei (AGNs), pulsars and supernova remnants (SNRs). A detailed description of the scientific goals of GLAST mission and an introduction to the experiment can be found in~\cite{GLAST01}.\\
The flight hardware production is now complete, the sixteen towers are integrated into the flight grid and the ACD is already placed over the LAT towers. The detector, which represents the biggest silicon strip tracker ever built, is taking data from cosmic rays, and will be integrated with the spacecraft within the next few months.

The scientific performance of the full LAT detector, i.e., the effective area, the point spread function and the energy dispersion, are obtained from detailed Monte Carlo studies, and are well-parameterized by a series of functions: the Instrument Response Function (IRF)\footnote{\url{http://www-glast.slac.stanford.edu/software/IS/glast_lat_performance.htm}}.

\section{Gamma-Ray Bursts and the Large Area Telescope: simulations and data analysis}

High energy emission from GRBs is still a puzzling topic and few observations are presently available above 50 MeV. EGRET detected only a few high-energy bursts \cite{Dingus95} and no apparent cut-off was detected at these energies. An extraordinary, and still unexplained, discovery was the delayed high energy emission\cite{Hurley94}, together with the more recent report of the observation of an additional high energy spectral component~\cite{Gonzalez03}.

For studying the performance of the LAT in observing GRBs, we set up a full simulation chain that starts from a detailed description of the sky, and adopts either a full Monte Carlo simulation of the detector (propagating every single particle through the different materials of the detector), or a fast science simulator which uses a parameterized description of the instrument for processing the incoming fluxes.

%\subsection{GRB models}

We have developed different GRB models within the framework of the LAT software\footnote{\url{http://www-glast.slac.stanford.edu/software/}}. The Gamma-Ray Burst physical model starts from the well known {\it fireball} model where the gamma rays are radiated by internal shocks \cite{Piran99}.
In this model, shells of matter are emitted with relativistic bulk Lorentz factor; faster shells overtake the slower ones and collide, producing internal shocks. In this scenario electrons and positrons are accelerated and lose energy via synchrotron radiation due in the presence of magnetic fields amplified by the shock compression. In this model the synchrotron cut-off at high energy due to the finite value of the Lorentz factor at which the electrons are accelerated is considered as well as the reprocessing of synchrotron radiation due to inverse Compton emission (i.e.,  a Synchrotron Self-Compton---SSC---spectrum). In this way we obtain a possible description of the high energy component based on ``physical'' assumptions. This model is suitable for studying the performance of the LAT detector in spectral analysis.
\begin{figure}
  \includegraphics[height=.3\textheight]{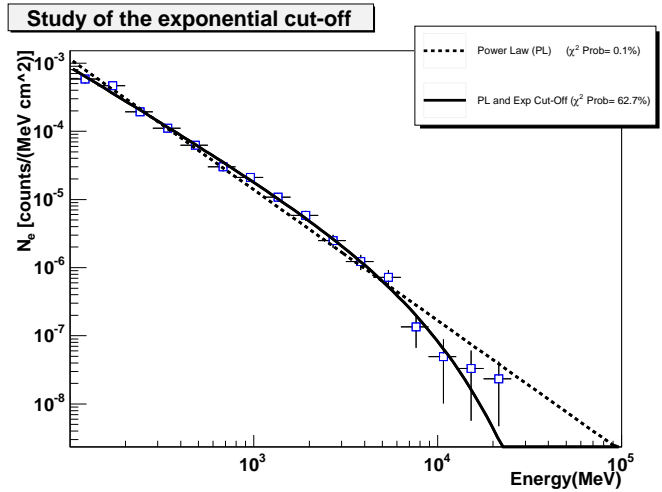}
  \caption{Study of the high energy cut-off. The simulated data have been produced using the GRB physical model with a high energy cut-off at 4.5 GeV. The reconstructed data are displayed in the plot and have been fitted with a power law model (dashed line) and with a power law model with an exponential cut-off (solid line). In the legend the values of the $\chi^2$ test are reported which denote the probability of obtaining a greater $\chi^2$ by chance. The estimated cut-off energy is $5.5\pm 1.5$ GeV.}\label{HECO}
\end{figure}
Fig.~\ref{HECO} shows the results of a simulation of a burst with a characteristic Lorentz factor $\Gamma=180$, which results in a synchrotron cut-off at 4.5 GeV. The large effective area and the good energy resolution of the LAT detector ($<$15\%) allows the study of the detectability of the high energy cut-off. 
The SSC emission operates in the opposite direction: enhancing the spectrum at high energy. 
This component has, to first approximation, in the same shape as the synchrotron component but, if $\gamma$ is the Lorentz factor of the incoming electron, is shifted by a factor of $\gamma^2\approx10^4$ to higher energy. Since the spectral energy distribution for the prompt  emission of a typical burst has a peak around a few hundred keV, the inverse Compton component, if present, is completely unobservable for detectors operating below tens of MeV (like BATSE); hence the search for the Compton component represents one of the key goals for the LAT detector onboard the GLAST satellite.
In addition, an energy-dependent time lag  resulting from Lorentz invariance breaking predicted by some quantum gravity theories\cite{amelino98,Omodei04QG} has also been included.

\begin{figure}
  \includegraphics[height=.45\textheight, angle=270]{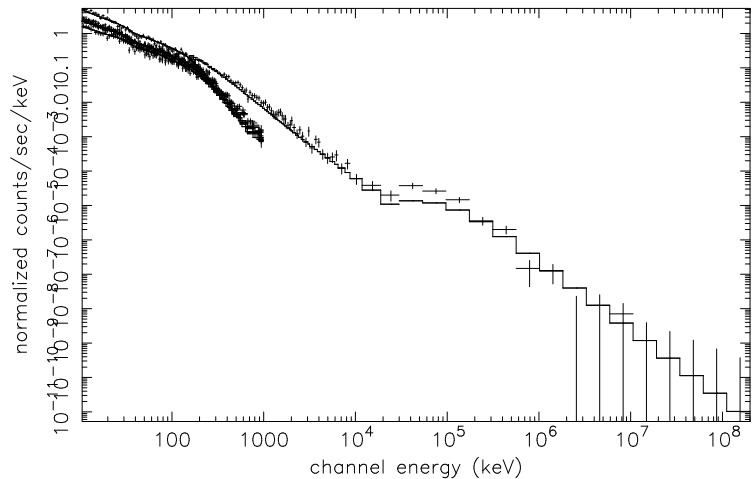}
  \caption{GBM (NaI and BGO) and LAT spectral fit. Producing this plot tested the full simulation  and  analysis chains. The output files are standard HEASARC FITS files, permitting  the spectral fit to be performed using XSPEC.} \label{GBMLAT}
\end{figure}

Our second approach to simulating GRBs in the LAT energy band, the GRB phenomenological model extrapolates a model of the GRBs spectrum and lightcurve in the $\sim100$ keV band to LAT energies.
For each simulated GRB the duration is drawn from the observed T$_{90}$  distribution and its fluence is sampled from the BATSE fluence distribution in the 50-300 keV energy range. Each burst has a spectrum described by the Band function\cite{Band93}, where the peak energy $E_p$ and the spectral indices $\alpha$ and $\beta$ are sampled from the observed distribution\cite{Band93, Preece00}.  The lightcurve is the sum of temporal pulses described by a general pulse equation with parameters drawn from observed distributions; the pulse width $W(E)$ scales with energy as $W(E)=W_0~E^{-0.33}$, as observed by Norris et al.\cite{Norris96}.
Short burst and long bursts are treated separately so that the observed hardness-duration correlation is reproduced. Analysis tools are needed for data handling and for science analysis.
Analyzing GRBs is different from analyzing other LAT sources. First, the GLAST Burst Monitor (GBM)\cite{GBM05} provides the soft gamma-ray counterparts to the LAT data; the GBM consists of 12 NaI detectors for the 10~keV to 1~MeV range and two BGO detectors for the 150~keV to 30~MeV range. Second, during the burst there are essentially no non-burst photons within the point spread function (PSF) of the LAT detector (3.5$^\circ$ at 100 MeV, 0.15$^\circ$ above 10~GeV). Finally, the Instrument Response Functions that are needed for converting the raw data into astronomical fluxes can be considered constant during the burst, contrary to stationary or long duration sources.
Fig.\ref{GBMLAT} shows an intense burst that has been simulated with  the full simulation chain. For this burst the LAT and GBM data have been simulated using the response functions for both types of detectors for a given the burst viewing angle. Within a few PSFs from the spatial barycenter we can assume that there are only burst photons for the LAT detector (i.e., we do not need a background model), while a background model has to be assumed for analyzing the GBM signal. The plot shows the result of a joint spectral analysis with one NaI detector (10 keV--1 MeV), one BGO detector (150 keV -- 30 MeV) and the LAT detector ($>$30~MeV).
GBM and LAT data will be jointly fitted providing spectral information over more than seven energy decades.
Here we estimate the LAT GRBs sensitivity,  the number of bursts detected per year as a function of the number of photons detected per burst. Of course this quantity depends on the GRB high energy emission where little information is available. In our computations we have adopted the phenomenological GRB description, providing a conservative description of the burst high energy flux.

\section{GRBs LAT sensitivity}

\begin{figure}[h!]
    \includegraphics[height=.33\textheight]{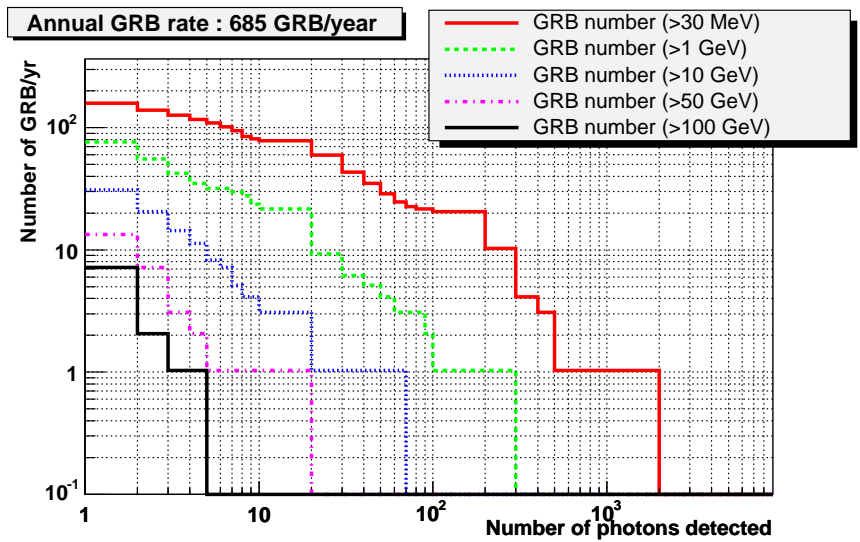}
  \caption{Model-dependent LAT GRB sensitivity assuming a mean burst rate of 650 bursts/yr, including the effect of the EBL absorption. Different curves refer to different energy thresholds.}
\label{Z}
\end{figure}

In order to compute the performance of the LAT detector to GRBs we simulate one year of observation in scanning mode, assuming a mean burst rate of 650 bursts per year full sky. Each burst is simulated with the phenomenological model discussed earlier.
An observed energy, an observed direction and a detection probability are computed for each simulated burst photon, taking into account the instrument response functions, resulting in an estimate of the number of photons that will be detected by the LAT detector. In addition, the orbit of the GLAST satellite, with SAA passages and Earth occultations, are considered.
At high energies ($>$10~GeV) it is important to consider the attenuation of the flux due to the cosmological absorption given by the interaction of a burst photon with an optical-UV photon of the extragalactic background light (EBL).
The uncertain EBL spectral energy distribution resulting from the absence of high redshift data provides a variety of theoretical models for such diffuse radiation.  Thus the observation of the high energy cut-off as a function of the GRB distance can, in principle, constrain the infrared background. In our simulation we have included this effect, adopting the EBL model proposed by  Primack\cite{Primack05} and assuming the long burst redshift distribution from Porciani and Madau\cite{PorcianiMadau01} and the short burst distribution from Guetta and Piran\cite{GuettaPiran05}.
We therefore plot the  number of expected bursts per year as a function of the number of photons per burst detected by the LAT (Fig.\ref{Z}). Different colors refer to different energy thresholds (see the legend).
The EBL attenuation affects only the high energy curve, as expected from the theory, leaving almost unchanged the sensitivities with thresholds less than 10 GeV. In this calculation, LAT will independently detect more then a hundred photons per burst for a few burst per year; these are the bursts for which a detailed spectral or even time resolved spectral analysis will be possible. Tens of bursts per month will result in more than ten counts in the LAT detector, and, with the assumed high energy emission model, a few bursts per year will show high energy prompt emission, with photons above 100~GeV.

\section{Alerts and Communication}
GLAST in the first year will operate in the scanning mode, providing uniform full sky coverage every three hours. Starting from the second year of operations, and depending on the Guest Investigator program, GLAST may  also be used in pointing mode.
The GBM will cover the entire visible sky not occulted by the Earth, and the LAT will cover 20\% the sky, with incident inclination up to almost $\sim$70$^{\circ}$ from the normal incident direction.
In addition GLAST will also be able to repoint in case of intense burst, to maintain the GRB in the LAT FoV during the prompt emission phase or to search for delayed emission. The GBM and LAT will  independently trigger on GRBs: the first on a rapid increase of the count rate, and the second considering spatial and temporal clustering of counts.  The GBM will detect $\sim$200 bursts per year, and more than 60 will fall in the LAT FoV, allowing joint observations. In case of a GRB trigger, an alert message will be send to the  ground via  TDRSS (a communications satellite system) within $\sim$10 seconds. This will provide basic information for follow up observations.  The initial on-board GBM localization accuracy is $\sim$15  degrees (within 1.8 s) that can be used by LAT. Updates will come later, reducing the GBM localization error box up to $\sim$5~degrees for a bright burst, while the LAT detector can provide better accuracy, up to tens of arcminutes (depending on the burst intensity). On-board LAT detection of GRBs is under development.
A full downlink of all the data will be performed via TDRSS  $\sim$~6 times a day and the scientific data, after a first analysis done by the LAT collaboration, will be delivered to the user community (for more details on GLAST operations in GRB observations, see\cite{McEnery}).

\section{Synergy with Swift}
A new mission will soon complement Swift's gamma-ray burst observations. The Gamma-ray Large Area Space Telescope (GLAST) will be launched in late-2007 and will cooperate with Swift in studying bursts. For a few bursts a year the LAT will localize bursts to sufficiently small error boxes that Swift can point to for follow up observations; for these bursts Swift will provide a more precise measurement of the GRB position. On the other hand, GLAST will frequently scan the burst position  in the hours after a Swift burst detection, searching for delayed high energy emission. 
Finally, simple computations show that$\sim$20 Swift-detected GRBs per year will also be in the LAT FoV.
%Swift and GLAST will provide for these bursts broad spectral coverage of the GRB spectrum (more than 7 decades, from 10 keV to hundreds of GeV) as well as afterglow observations and possibly a burst redshift, with the resulting impact on the GRB science.

%\bibliographystyle{aipproc}
%\bibliography{Omodei}
%\IfFileExists{\jobname.bbl}{}
% {\typeout{}
%  \typeout{******************************************}
%  \typeout{** Please run "bibtex \jobname" to optain}
%  \typeout{** the bibliography and then re-run LaTeX}
%  \typeout{** twice to fix the references!}
%  \typeout{******************************************}
%  \typeout{}
% }

\end{document}